\documentclass[pra,twocolumn,superscriptaddress]{revtex4}
\usepackage{color}
\usepackage{amsfonts}
\usepackage{amsmath}
\usepackage{amssymb}
\usepackage{graphicx}
\usepackage{empheq}
\usepackage{subfigure}
\usepackage[usenames,dvipsnames]{xcolor}
\usepackage{soul}
\usepackage[title]{appendix}

\begin{document}

\title{Edge states in a two-dimensional non-symmorphic semimetal}

\author{P.G. Matveeva}
\affiliation{Department of Physics and Research Center Optimas, University of Kaiserslautern, 67663 Kaiserslautern, Germany}
\author{D.N. Aristov}
\affiliation{``PNPI'' NRC ``Kurchatov Institute'', Gatchina 188300, Russia}
\affiliation{Department of Physics, St.Petersburg State University, Ulianovskaya 1,
St.Petersburg 198504, Russia}
\affiliation{Institute for Nanotechnology, Karlsruhe Institute of Technology, 76021
Karlsruhe, Germany }
\author{D.  Meidan}
\affiliation{Department of Physics, Ben-Gurion University of the Negev, Beer-Sheva 84105, Israel}
\author{D.B. Gutman}
\affiliation{Department of Physics, Bar-Ilan University, Ramat Gan, 52900, Israel}

\begin{abstract}
Dirac materials have unique transport properties, partly due to the presence of surface states. A new type of Dirac materials, protected by non-symmorphic symmetries was recently proposed by Young and Kane \cite{KaneYoung}. By breaking of time reversal or inversion symmetry one can split the Dirac cones into Weyl nodes. The later are characterized by local Chern numbers, that makes them two-dimensional analogs of Weyl semimetals. We find that the formation of the Weyl nodes is accompanied by an emergence of one-dimensional surface states, similar to  Fermi arcs in Weyl semimetals and 
edge states in two-dimensional graphene. We explore these  states for a quasi-one-dimensional non-symmorphic ribbon. The type and strength of applied deformation control the location and 
 Weyl nodes and their composition.  This  determines  the properties of  emerging  edge states. The sensitivity of these edge states to the external deformations makes non-symmorphic materials potentially useful as a new type of electromechanical sensors.   
\end{abstract}
\date{\today}
\pacs{}
\preprint{}

\maketitle

\section{Introduction} 
Topological phases of matter are attracting a growing attention in recent years. Due to their unique band structure, these gapped phases are known to support protected surface states. 
Weyl semimetals \cite{Binghai1,Xu,Lv,Lu,Burkov} are not 
topological in a strict sense, as they do not possess a non-zero Chern number, yet some of their properties have topological origins. 
This happens due to the special points in a Brillouin zone, known as Weyl nodes, that can be attributed to a local Chern number. Their presence gives rise to a number of interesting effects associated, for example, with a chiral anomaly \cite{NielsenNinomiya, Burkov2}. 
 One of the striking features of the Weyl semimetals is an emergence of the Fermi arcs, two-dimensional surface states that are robust against the disorder and a surface preparation. The position of the Fermi arcs in the $k$-space is unambiguously determined by the position of the Weyl nodes by a projection of these point onto a surface\cite{Wan}. Such states were numerically predicted in different classes of realistic materials \cite{Wan,Weng,Singh, Potter,Huang,Solyanov,Binghai2,Binghai3} and were experimentally  observed  via angle-resolved photoemission spectroscopy \cite{Xu,Deng,Li,Jiang,Batabyal}.

In three dimensional materials the Weyl nodes are stable \cite{Armitage} and can not be removed by any (sufficiently weak) perturbation.
The situation in two dimensions is more fragile as robust  Weyl semimetals in two dimension do not exist. 
The closest analog was recently proposed in \cite{KaneYoung}, who considered two-dimensional lattice 
with non-symmorphic symmetries, that combine half-translations with mirror reflections or rotations. 
By breaking the time reversal (TR) and inversion symmetries one  splits the Dirac points into Weyl nodes. 
The stability of Weyl nodes (against opening a gap) is protected by a non-symmorphic lattice symmetries  \cite{Symmetries_in_solids}. The spectrum of an emergent two dimensional Weyl semimetal is somewhat similar to graphene. It exhibits an even number of band touching points with Dirac type singularity, similar to $K$ and $K'$ points in graphene. Near these points the Berry curvature has  $1/k^2$ singularity  and they are associated with a local Chern number $\pm 1$. 

The presence of the Dirac cones in graphene leads  the formation of edge states \cite{FujitaJPSJ1996,NakadaPRB1996,WakabayashiPRB1999}.
For finite graphene ribbons, an edge mode emerges  between the projections of the $K$ and $K'$ points on the direction of the boundary. This yields the longest (in a $k$-space) edge in a zig-zag type of boundary, that smoothly shrinks as the direction of the boundary changes, while  vanishing completely  for an armchair boundary, where  the points  $K$ and $K'$  are projected on top of each other. In this case the edge state has a zero length, i.e. disappears. 
For a semi-infinite graphene ribbon in the presence of the chiral symmetry, the edge modes are flat. The breaking of this symmetry turns graphene into  an insulator (trivial or topological)\cite{Kane_Mele} and induces a dispersion of the edge state.

The stability of the Dirac points in graphene  has topological origins. For spinless graphene the existence of Dirac nodes can be proven by  computing the flux of the Berry curvature through the half of the Brillouin zone\cite{Kane_Fu}. As a consequence, while the distortion  of graphene layer shifts the positions of the Dirac points in the  $k$-space\cite{Yang},  no gap opens and the spectrum near the degeneracy points has Dirac type singularity. Note, that for the unstrained pristine graphene the degeneracy points are located at the high symmetry points. The spectrum degeneracy at these points  is  determined by crystal symmetry and   follows from the two-dimensional irreducible representation of the little group.  For the distorted graphene, on the other hand,  the position of the degeneracy points is generic and does not correspond to  any high symmetry point. 
 However, their existence is guaranteed by the topological arguments, that allow adiabatically connect the "distorted" graphene Hamiltonian with Dirac point "somewhere"  in the Brillouin zone to the one of ideal graphene, with the edge state following this evolution. Therefore there is a  connection between  the edge state in unstrained graphene    
and its topological properties. This connection is well established for a system 
with chiral symmetries\cite{Ryu}.  Our work suggests that this proof can be further generalized to the
situations where this symmetry is weakly violated.

In this manuscript, we study the formation of the edge states in 2D lattice models with non-symmorphic symmetries.
We consider the model put forward in Ref.  \cite{KaneYoung}. The spectrum has  three Dirac nodes,  
protected by crystal non-symmorphic symmetries, inversion and time reversal (TR) symmetries. The  reduction of some of these symmetries splits the  Dirac nodes into the Weyl nodes \cite{KaneYoung}. We show here that this is inevitably accompanied  by the formation  of  edge states. However, the nature of these edge states, such as their spin polarization, depends on the details of the applied deformation.

The mere presence of Dirac type singularities in the spectrum leads to the emergence of the edge states.
Such systems fall into a category of distorted graphene Hamiltonians and consequently support   edge states. Of course, this approximation overlooks the non-singular parts of the spectrum,
that may lead to the additional surface states\cite{Levitov}.
Moreover, if this Dirac type and the regular states inhabit the same part of the phase space, the hybridization of this mode near the boundary can occur with an unknown outcome \cite{Kharitonov}. 
We note that  in the model we consider, there are no regular parts of the spectrum in the vicinity of the  Weyl nodes.

The paper is organized as follows. In section II we rederive the results of Ref.\cite{KaneYoung} with some more details. In section III we study a formation of the Weyl nodes protected by non-symmorphic symmetries and corresponding edge states in a finite ribbon. In section IV we construct perturbation that breaks non-symmorphic symmetries yet "accidentally" leads to Weyl nodes. We conclude by discussing the properties of the edge states and their potential applications.

\section{Dirac semimetal in 2D}
\label{Dirac_in2d}
\begin{figure}
\centering
\includegraphics[width=1\linewidth]{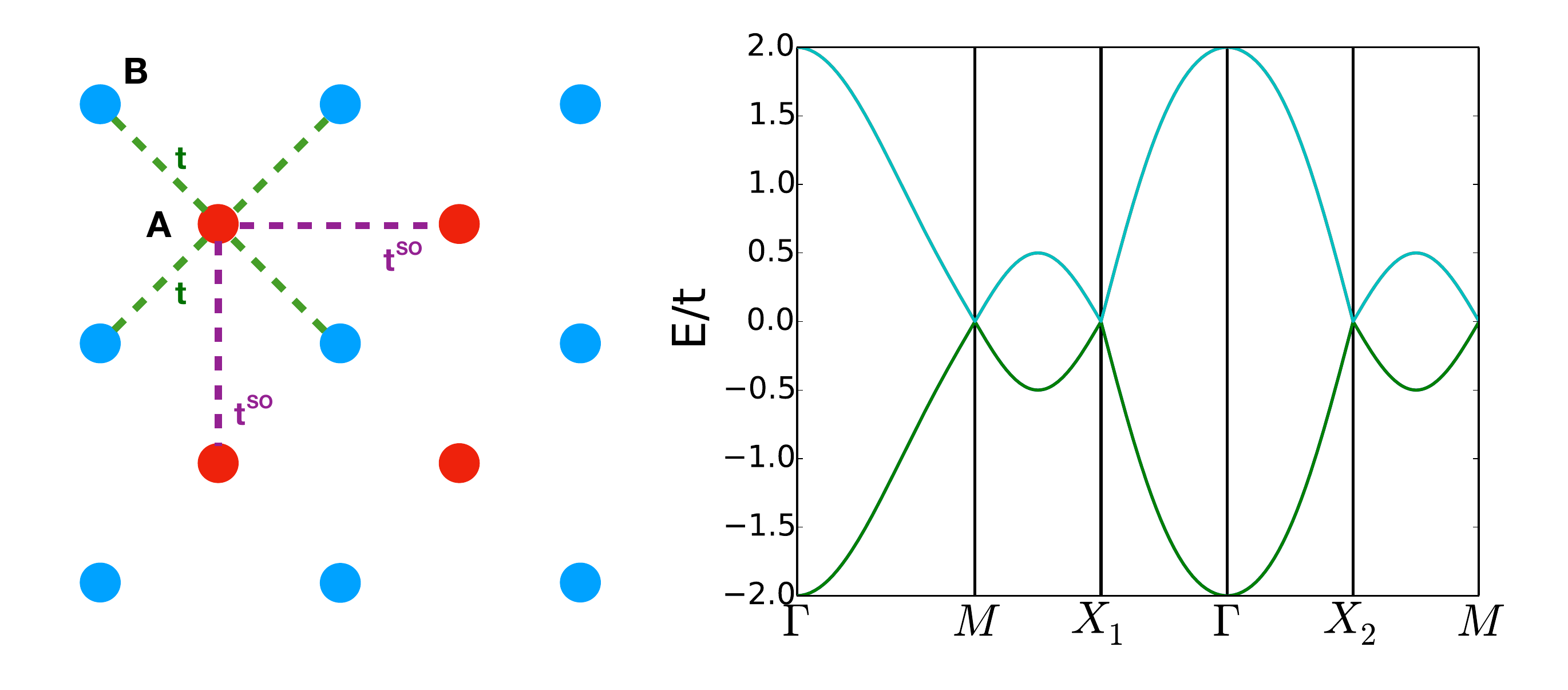}
\caption{\label{lattice}
Left panel: The schematic lattice. Right panel: Spectrum  for Hamiltonian   Eq.(\ref{h0}) with $t=1,t^{\rm SO}=0.5$. In the presence of the time reversal symmetry each line is two  times degenerate. 
High  symmetry points  $\Gamma= \{0,0\}, M=\{\pi,\pi\}, X_1=\{\pi,0\}$, $X_2=\{0,\pi\}$. }
\end{figure}

We focus on the model proposed in \cite{KaneYoung}.  In order to keep the presentation self-contained   
we briefly repeat their findings and explain the relationship between the symmetry of the system and the spectrum degeneracy.   
In 2D, the simplest lattice with a non-symmorphic symmetry is shown in Fig.\ref{lattice}. 
It  is a square lattice with two atoms in the unit cell. One of the atoms is shifted out of the plane  (along $\hat{z}$ direction). 
It is accounted by the layer group P4/nmm.
Within the tight-binding approximation the Hamiltonian is given by
\begin{flalign}
\label{h0}
H_0 =& 
2t\tau_x\cos\frac{k_x}{2}\cos\frac{k_y}{2} + t^{\rm SO}\tau_z[\sigma_y\sin k_x-\sigma_x\sin k_y ].
\end{flalign}
Here $\tau$ and $\sigma$ are Pauli matrices in the sublattice and spin space,
$t$ is an amplitude of nearest neighbor hopping and $ t^{\rm SO}$ is an amplitude of the next nearest neighbor spin-orbit interaction. 
The system posses  an  inversion symmetry, represented in  the  momentum space by  $\mathcal{P}=\{\tau_x| \mathbf{k} \rightarrow -\mathbf{k} \}$. It also preserves  time-reversal symmetry 
$\mathcal{T}=\{ i\sigma_yK| \mathbf{k}\rightarrow -\mathbf{k}\}$, $\mathcal{T}^2=-1$. In addition the model respects the non-symmorphic symmetries that combine a translation 
by a half of the lattice constant with rotation and mirror reflections.
To account for these symmetries we define the operators $\hat{t}_x$ and $\hat{t}_y$  that describe half-translations along the axes $\hat{x}$ and $\hat{y}$.
They act on the Bloch states as
$\hat{t}_x|\Psi_n\rangle=e^{i\frac{k_x}{2}}|\Psi\rangle $, $\hat{t}_y|\Psi_n\rangle=e^{i\frac{k_y}{2}}|\Psi_n\rangle$, where $n$ the is a band number. 
The rotation by $\pi$ around the $x$ axes acts on  the orbital part of the wave function as $\hat{S}_x(k_x,k_y)=(k_x,-k_y), \text{i.e.} k_y \rightarrow -k_y$ (and similarly for the rotation around $y$-axes).  
In terms of these operators the three non-symmorphic symmetries can be written as follows:
\begin{enumerate}
\item
Rotation around $\hat{x}$ axis together with a half-translation along this axis\\
$g_1=\{C_{2x}|\frac{1}{2}0\}=i\sigma_x\tau_x \hat{S}_x \hat{t}_x$. 
\item Rotation around $\hat{y}$ axis together with a half-translation along this axis: \\
$g_2=\{C_{2y}|0\frac{1}{2}\} =i\sigma_y\tau_x \hat{S_y}\hat{t}_y$  
 \item Mirror reflection around $z$ plane together with two half-translations:\\
 $g_3=\{M_{z}|\frac{1}{2}\frac{1}{2}\}=i\sigma_z \tau_x \hat{t}_x \hat{t}_y$
\end{enumerate}

Band crossing must happen at any  $g_i$--invariant line.
Indeed, on this line the Hamiltonian and the corresponding symmetry operator can be diagonalized simultaneously, 
so  $g_i |\Psi^{\pm}_{n}(\mathbf{k}) \rangle=\pm \lambda e^{i \mathbf{k}\mathbf{t}} |\Psi^{\pm}_{n}(\mathbf{k}) \rangle $.
Since  $e^{ i\mathbf{\mathbf{G}}\mathbf{t}}=-1$,  the two eigenstates switch places as one moves from the point  ${\bf k}$  in the Brillouin zone to  the  equivalent point $\mathbf{k} \rightarrow \mathbf{k}+\mathbf{G}$ along the $g_i$ invariant line.
That means that the two eigenstates must switch places an odd number of times, generating an odd number of crossing  points on this  line  in the Brillouin zone.

As long as  the  time-reversal and inversion symmetry are preserved,  the spectrum is doubly degenerate.  
Since the point, $\mathbf{k}=\mathbf{G}/2$ remains invariant under the reversal  of time    it must be a crossings point.
Moreover, the spectrum near this point must be symmetric and  any of the crossing points of the spectrum is fourfold degenerate.
Therefore  in the presence of  time reversal, inversion and  any non-symmorphic  symmetries Dirac semimetal is formed. 
The arguments so far  were  quite general and they hold for any lattice consistent with these symmetries.
Now we focus  on the  lattice  shown in Fig.\ref{lattice}. It has the following $g_i$ invariant lines in the Brillouin zone:  $g_1$ invariant lines  $k_y=0, \pm \pi$ ,  $g_2$ invariant lines $k_x=0, \pm \pi$.
The  Dirac points  $X_1=\{\pi,0\}$, $ X_2=\{0,\pi\}$  
and $M=\{\pi,\pi\}$ lie on the intersection of those lines.

The Hamiltonian may be reduced to a block diagonal form  by the unitary transformation 
\begin{equation}
\label{unitary_transformation}
U=\left(\begin{matrix}
\sigma_0 & \sigma_x \\
\sigma_z  & -i\sigma_y
\end{matrix} \right)
\end{equation}
In this basis the Hamiltonian $H_0$ reads 
\begin{equation}
\label{blockhamiltonian}
\tilde{H}=U^{-1}HU=\left(\begin{matrix}
H_{-} & 0\\
0 & H_{+}
\end{matrix} \right)\,.
\end{equation}
The diagonal elements 
\begin{equation}
\label{blocks}
H_{-}=\mathbf{d}\cdot \boldsymbol{\sigma^*}\,, \,\,\,\,\,\,
H_{+}=\mathbf{d}\cdot \boldsymbol{\sigma}\,.
\end{equation}
and 
\begin{equation}
\label{dvector}
\mathbf{d}=\left \{-t^{\rm SO}\sin k_y, -t^{\rm SO}\sin k_x, 2t\cos \frac{k_x}{2}\cos\frac{k_y}{2}\right \}.
\end{equation}
Close to the Dirac points $M, X_1$ and $X_2$ the Hamiltonian (\ref{blockhamiltonian}) is mathematically equivalent to a graphene-like Hamiltonian \cite{CastroNeto}.
Indeed, by  expanding   the Hamiltonian (\ref{blockhamiltonian})  near the Dirac points up  to linear order in ${\bf k}$ 
and performing an additional unitary transformation one finds
\begin{equation}
\label{M_point}
\tilde{H}=-t^{\rm SO}\left(\begin{matrix}
\boldsymbol{\sigma} \cdot {\bf k} & 0\\
0 & \boldsymbol{\sigma^*} \cdot {\bf k}
\end{matrix} \right)\,.
\end{equation}

The resulting Hamiltonian (\ref{M_point}) is formally equivalent to the  Hamiltonian of spinless graphene 
in the vicinity of $K$ and $K'$ points in graphene, with coinciding  $K$ and $K'$ points.
Using the continuous  description,  one can  analyse the change of the spectrum in the presence of 
a generic perturbation:
\begin{small}
\begin{equation}
\label{blockhamiltonian_pert}
\frac{H}{-t^{\rm SO}}=\left(\begin{matrix}
\boldsymbol{\sigma} \cdot({\mathbf {k}}-{\mathbf {a}}) +m_1(\mathbf{k})\sigma_z& \hat{m}({\mathbf {k}})\\
\hat{m}^*({\mathbf {k}}) & \boldsymbol{\sigma^*} \cdot({\mathbf {k}}+{\mathbf{a}}) +m_2(\mathbf{k})\sigma_z
\end{matrix} \right)\,
\end{equation}
\end{small}
with $\hat{m}({\mathbf k})=\mathbf{m}\boldsymbol{\sigma}+m_0\sigma_0$.

We first consider the case when the terms that open a gap are absent, i.e. $m_1(\mathbf{k})=m_2(\mathbf{k})=0$.  
 Keeping   ${\bf m} \equiv (m_x,m_y,m_z) =(0,0,m_z)$, 
 both the vector ${\bf a}$ and  $m_z,m_0$ account for the  splitting of the Dirac node at ${\bf k}=0$ into two Weyl nodes located  at $\mathbf{k}_{\pm}=\left\{\text{Re} \left[C\right], \pm \text{Im} \left[C \right] \right\}$, with $ C=i \sqrt{(a_x+ia_y)^2+m_z^2-m_0^2}$.
 When these perturbations are absent, $a=0,m_0=0, m_z=0 $, but  $m_x \neq 0, m_y \neq 0$,  nodal lines emerge. The  position of these lines  in $k$ space is determined  by the equations $k=|m_x+m_y|, k=|m_x-m_y|$, which  trace an ellipse in $k$ space. 
 For  $\mathbf{m}_{1,2}({\bf k})\neq 0$ a  gap opens, unless 
  $\mathbf {m}_{1,2}({\bf  k}) = 0$ vanishes either at  $\mathbf{k}_{\pm}$
  or at some point on the nodal lines, where a Weyl node emerges. 
  Due to the Nielsen-Ninomiya theorem, the number of emergent Weyl nodes must be even\cite{NielsenNinomya2}.
 In terms of our model  (\ref{blockhamiltonian_pert}), this implies that  the functions  
 $m_1(\mathbf{k})$ and $m_2(\mathbf{k})$ are not independent,  but  vanish in  a way that an even number of Weyl node arises and the total topological charge remains zero. 
 Notice also, that in (\ref{blockhamiltonian_pert}) we do not consider the perturbations that proportional to $\sigma_z \otimes I_2$: even though they lift the degeneracy of the Dirac point, the resulting nodes with the opposite topological charge remain in the same point in $k$-space. 
 
In the presence of  non-symmorphic symmetries the functions $m_1(\mathbf{k})$ and $m_2(\mathbf{k})$ vanish in at least one point along  the corresponding high symmetry line\cite{KaneYoung}.
Therefore the formation of Weyl nodes, in this case, is guaranteed by the symmetry.
We will refer to such points symmetry protected Weyl nodes.
We name  Weyl nodes accidental if their emergence cannot  be  established solely on the basis of 
 lattice symmetries.

\begin{figure}
\centering
\includegraphics[width=0.6\linewidth]
{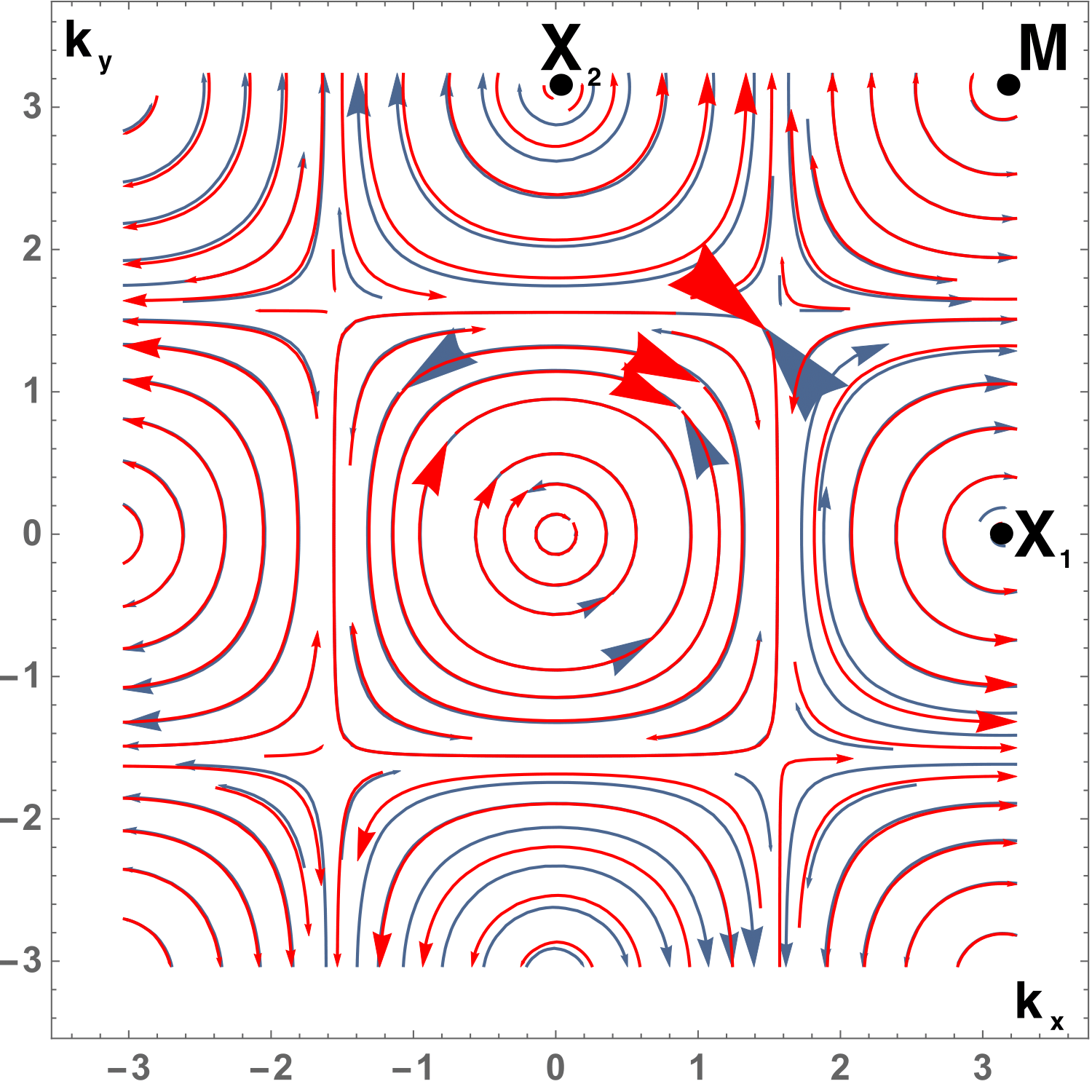}
\caption{\label{Berry_Dirac}
Berry connection for the model $H_0$. It illustrates, that the total Berry connection and correspondingly the topological charge at each point in the Brillouin zone are zero.
Clockwise and counterclockwise directions correspond to the connections of the opposite signs.}
\end{figure}

We next consider the case when Weyl nodes emerge.
Since they have the opposite Chern numbers, the total Chern number at the Dirac point is zero.
Indeed,  the Berry flux through a cross-section $\mathcal{S}$ in the k-space is given by an  integral of the Berry curvature 

\begin{equation}
\label{Berry_phase}
\phi_n=\int\limits_{\mathcal{S}} \mathbf{B}_{n}(\mathbf{k}) \cdot \text{d}\mathbf{S},
\end{equation}
where $n$--band number. Berry curvature is defined as the curl of the Berry connection $\mathbf{A}_{n}(\mathbf{k})$:
\begin{equation}
\label{berrycurvature}
\mathbf{B}_{n}(\mathbf{k})=\nabla_{\mathbf{k}} \times \mathbf{A}_{n}(\mathbf{k}),
\end{equation}
where the Berry connection of the $n$th  band is given by 
\begin{equation}
\label{berryconnection}
\mathbf{A}_{n}(\mathbf{k})= i\langle \Psi_n| \nabla_{\mathbf{k}} |\Psi_n \rangle.
\end{equation}
In two dimensions Berry curvature has only one component
\begin{equation}
\label{berry_curvature_2D}
B_{n}(\mathbf{k})=-2\text{Im} \langle \partial_{k_x} \Psi_n |\partial_{k_y} \Psi_n \rangle.
\end{equation}
The Berry connection for Eq.(\ref{h0}) is shown in  the Fig. \ref{Berry_Dirac}. 
It demonstrates, that the field $\mathbf{A}_{n}(\mathbf{k})$ has the opposite sign for different $n$ ($\mathbf{A}_{1}(\mathbf{k})=-\mathbf{A}_{2}(\mathbf{k})$). Therefore the  total Berry curvature and phase vanish, $B(\mathbf{k})=\sum_n B_n(\mathbf{k})=0$, $\phi =\sum_n \phi_n=0$, everywhere in the Brillouin zone. 

We next discuss the lowering of degeneracy at the Dirac node which splits it into two Weyl nodes.
As is well known \cite{Armitage}, the existence of the Weyl nodes requires violation of time reversal or inversion symmetry (or both).  However, unlike in the three-dimensional case, where a Weyl node is stable against any (sufficiently weak)  perturbation,  the degeneracy in $d=2$ can be completely lifted by a generic perturbation.
To guarantee that the spectrum remains gapless one has to keep at least one of non-symmorphic symmetries intact.   We now discuss several representative examples of such perturbations.
One of them is consistent with the discussed symmetry-related arguments, and another one results in the accidental Weyl points. In the latter case, the emergence of the Weyl nodes is not guaranteed  by the symmetry arguments and happens accidentally.
In both cases, we show that the emergence of Weyl nodes is accompanied by the edge states.

\section{Symmetry protected Weyl nodes}
\label{symmorphic}
We now consider a perturbation that breaks  time-reversal symmetry.
It arises if  one of the atoms in the elementary cell is displaced in the $\hat{y}$ direction and additionally has a magnetic dipole moment alligned in $\hat{x}$ direction, thus resulting in the spin-dependent nearest-neighbor hopping amplitude. It breaks time-reversal symmetry, but not any of non-symmorphic:
\begin{equation}
\label{pert1}
V_1=v_1\tau_y \sigma_x \sin \frac{k_y}{2} \cos \frac{k_x}{2}\,.
\end{equation}
We mention in passing that the spin-dependent hopping in this case happens even in the absence of the lattice deformation, but the resulting term of the Hamiltonian 
\begin{equation}
\label{spin_hopp}
V_{\text{hop}}=t_2\tau_x \sigma_x \cos\frac{k_y}{2} \cos \frac{k_x}{2}
\end{equation}
does not affect the splitting of Weyl nodes. This is because it  vanishes  along $g_i$ invariant  lines ($k_x= \pm \pi, k_y=\pm \pi$), where the band crossing happens. 

\begin{figure}
\centering
\includegraphics[width=\linewidth]{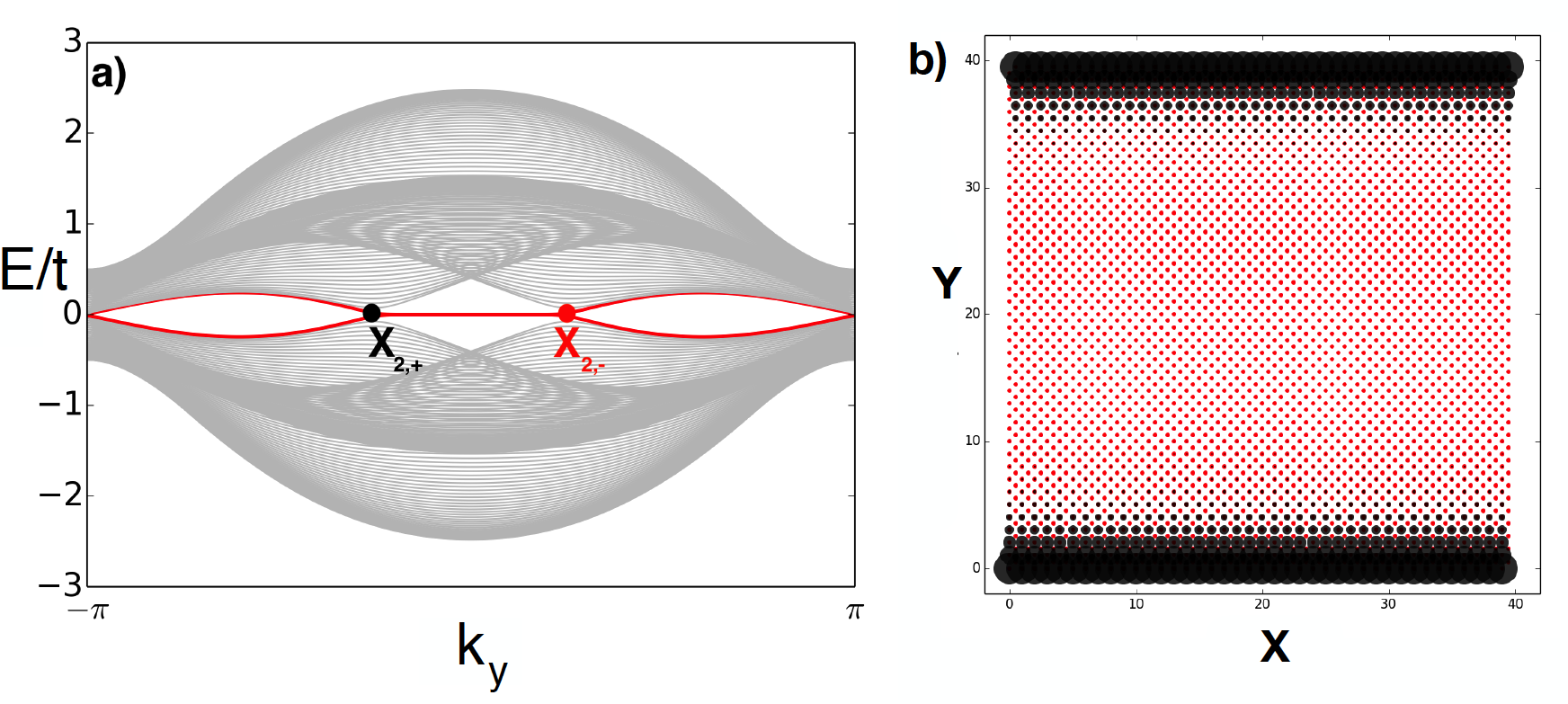}
\caption{\label{Perturbation1_edge}
a) Spectrum of the model $H_0+V_1+V_{\text{hop}}$ for the ribbon with the width $L_y=60$, red line represents the band with the dispersionless edge states that exist between the Weyl points $X_{2+}$ and $X_{2-}$ b) Distribution of the amplitude of the wavefunction in the ribbon with the width $L_y=40$ for the edge state with $k_x=0$}
\end{figure}

Due to the perturbation (\ref{pert1}) the Dirac node $X_2$ splits, as shown in Fig.  \ref{Perturbation1_edge}. The resulting  Weyl nodes are located at the line $k_y=\pi$, which is protected by $g_1=\{C_{2x}|\frac{1}{2}0\}$. 
The two Weyl points $X_{2+}$ and $X_{2-}$ have the coordinates:
 \begin{equation}
 \label{weyl_points_pert1}
 X_{2\pm}=\left(\pm 2\arcsin \left[v_1/2t^{\rm SO} \right] ,\pi\right) 
 \end{equation}
In the bulk spectrum on the $g_1$ invariant  line is continuous and is given by 
 \begin{equation}
 \label{spectrum_pert1}
\begin{split}
\epsilon_{1,2}= \pm t^{\rm SO} \left[ \frac{v_1}{t^{\rm SO}}\cos \frac{k_x}{2}-\sin k_x \right] \\
\epsilon_{3,4}= \pm t^{\rm SO} \left[ \frac{v_1}{t^{\rm SO}}\cos \frac{k_x}{2}+\sin k_x \right] .
\end{split} 
 \end{equation}
We now focus on the edge states.  Following the discussion above, we expect the edge state to form between the projection of Weyl  nodes on momentum axes parallel to the boundary.
For a small perturbation $v_1/t^{\rm SO} \ll 1$ the distance between Weyl points (and the length of the edge state in the k-space) scales linearly with the perturbation: 
 \begin{equation}
 \label{weylp_dist_pert1}
 \Delta k_x =\frac{2v_1}{t^{\rm SO}}
 \end{equation}

Fig. \ref{Perturbation1_edge} shows the spectrum of the model given by $H_0 + V_1+V_{\rm hop} $ for a ribbon infinite in $\hat{x}$ direction and with  a finite width $L_y$ in $\hat{y}$ direction.
The results show the appearance of a flat band  edge state confined to the boundaries in the real space and connecting  the projections of $X_{2,+}$ and $X_{2,-}$ points in $k$-space.  
 \begin{figure}
\centering
 \includegraphics[width=0.6\linewidth]{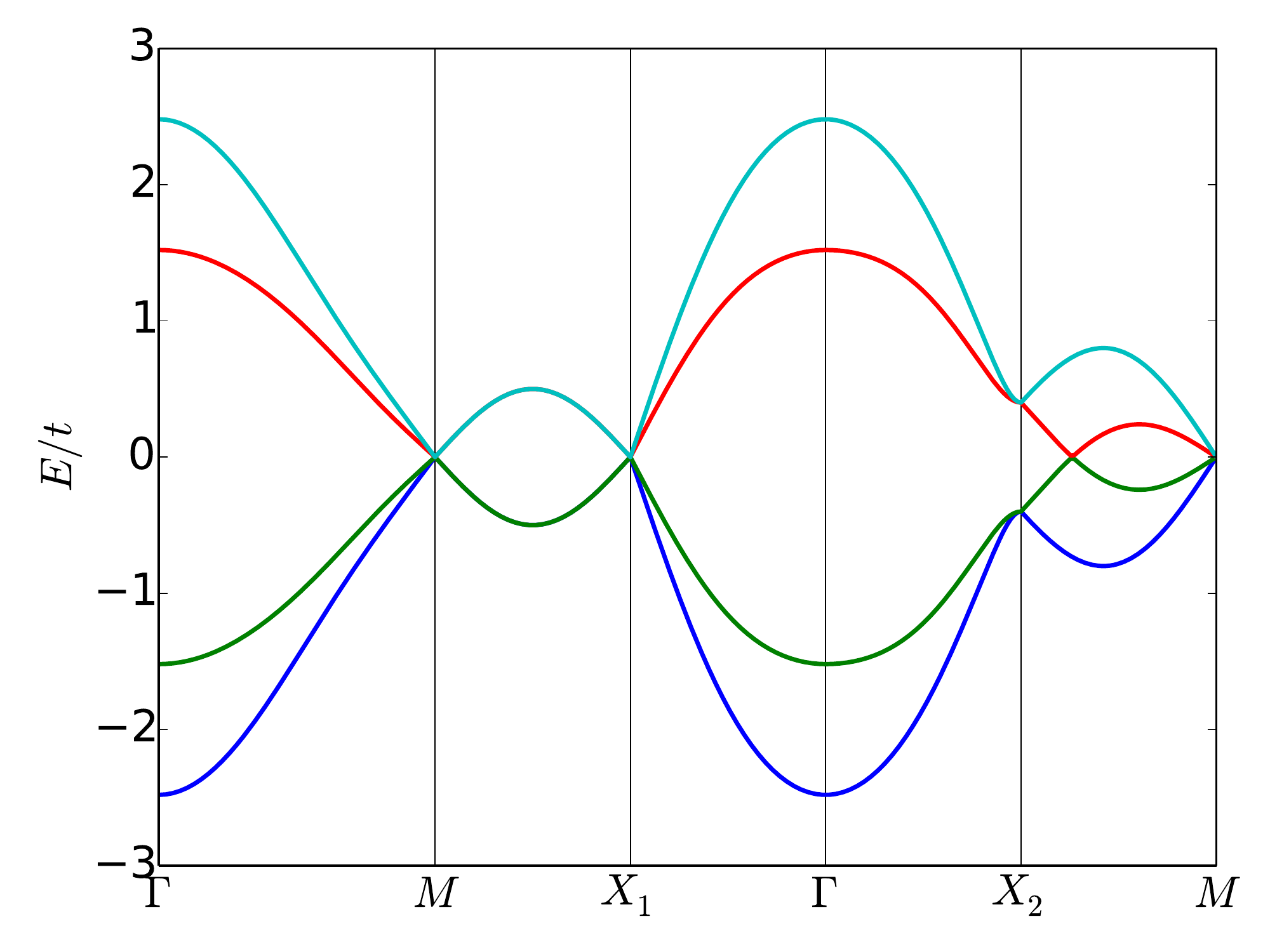} \\
 \includegraphics[width=0.6\linewidth]{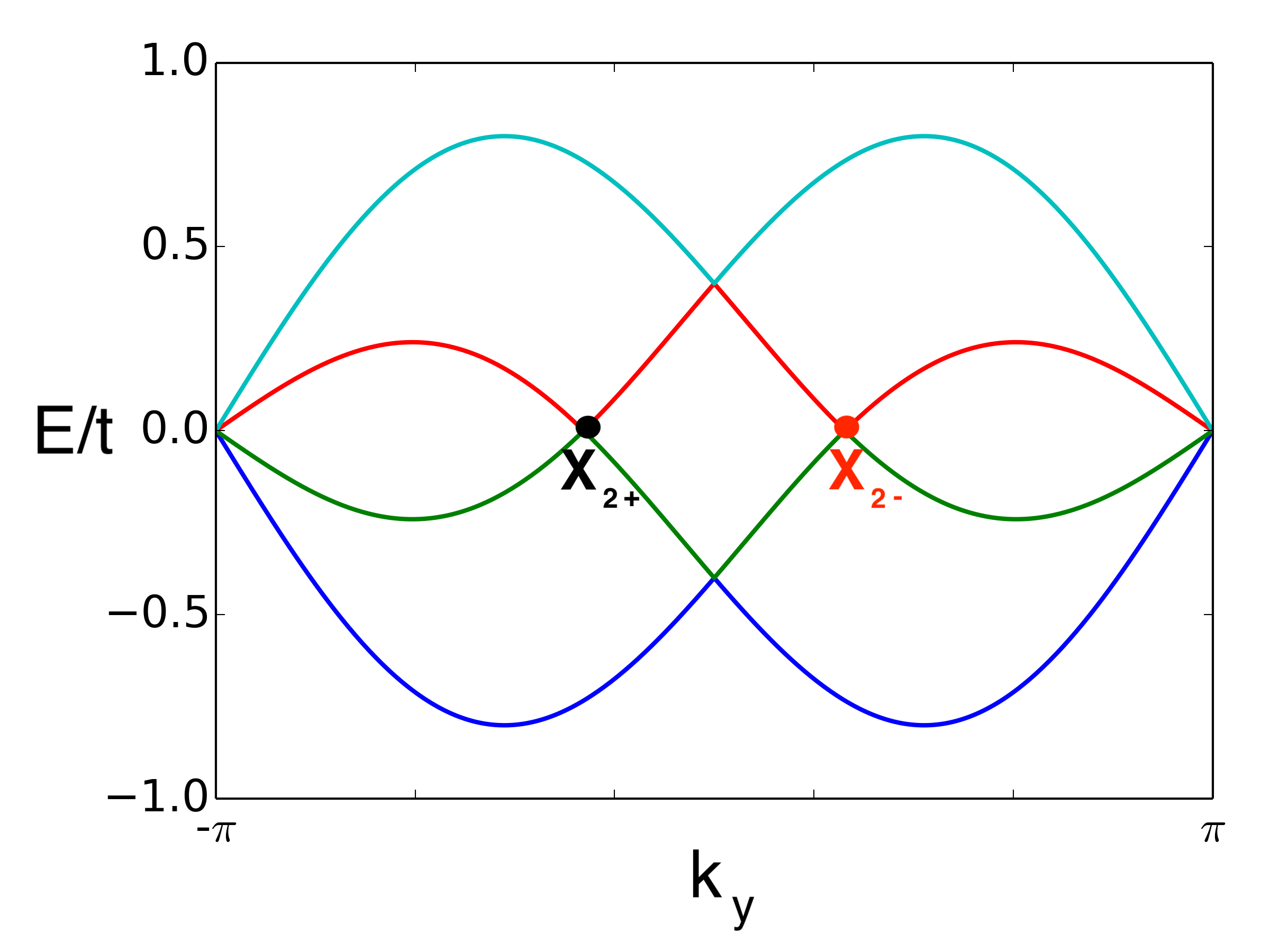}
\caption{\label{Perturbation1_bulk}
Spectrum of the model $H_0+V_1+V_{\text{hop}}$.  Used parameters (in the units of $t$) are: $t^{\rm SO}=0.5, t_2=0.12, v_1=0.1$. The perturbation splits the Dirac point $X_2$ into two Weyl nodes $X_{2+}$ and $X_{2-}$. }
\end{figure}
We now proceed to  a case where non-symmorpic  symmetries are broken and the existence of Weyl nodes is not determined by symmetry.

\section{Accidental Weyl nodes \label{accidental}}

\begin{figure}
\centering
\subfigure{
\includegraphics[width=0.6\linewidth]{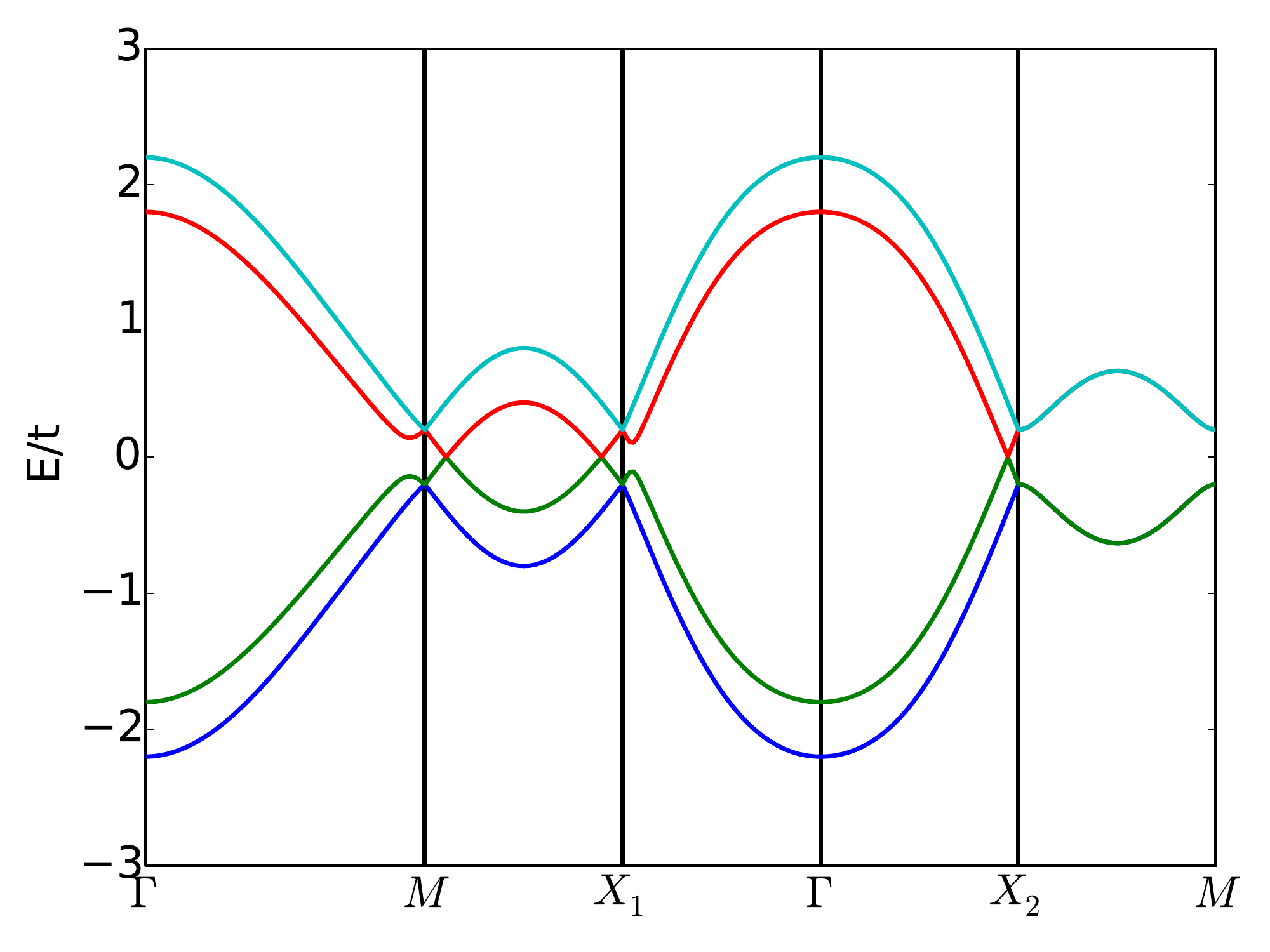}}
\subfigure{
\includegraphics[width=0.6\linewidth]{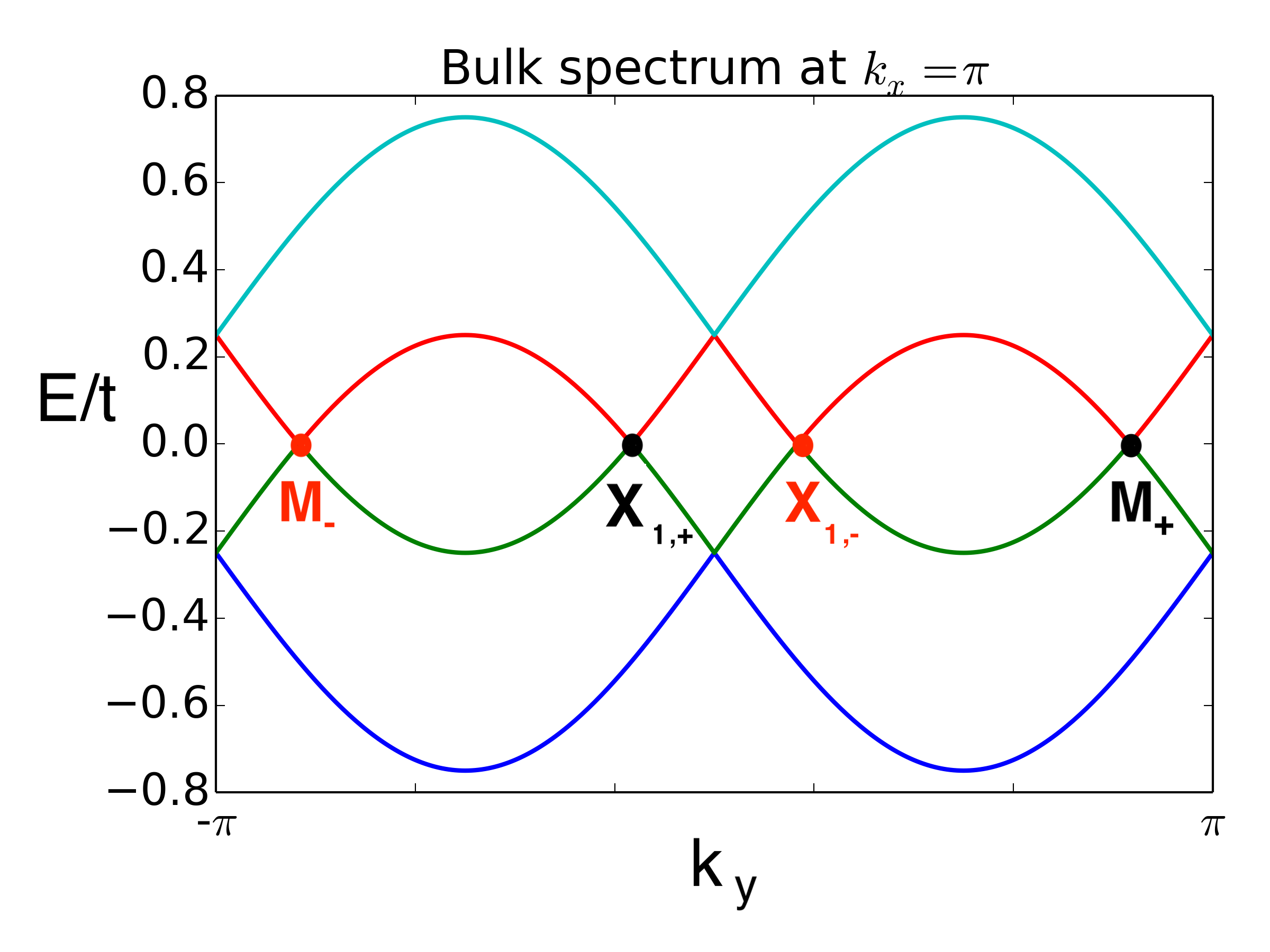}
}
\subfigure{
\includegraphics[width=0.6\linewidth]{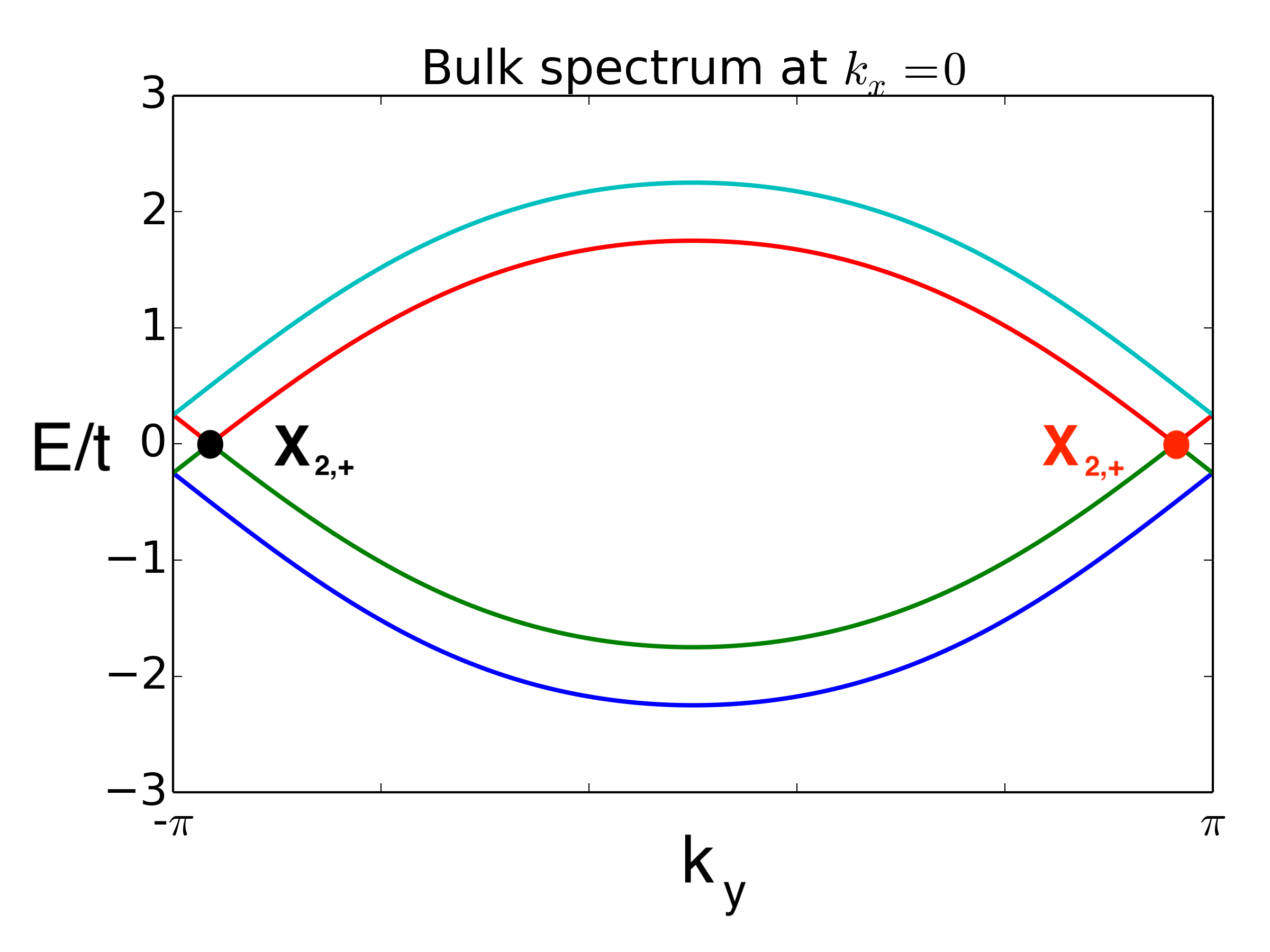}
}
\caption{\label{Magnetic_field_bulk} Bulk spectrum of the model $H_0+V$. Used parameters (in the units of $t$) are: $t^{\rm SO}=0.3, B=0.2$}
\end{figure}

In this section we study the effect of external in-plane magnetic field coupled to the spins of electrons. The perturbation is described by the following term 
\begin{equation}
\label{magn_field}
V=B_x\sigma_x,
\end{equation}
where the magnetic field $B$ points in $\hat{x}$ direction. 
This perturbation breaks time-reversal symmetry $\mathcal{T}$, non-symmorphic symmetries $g_2$, $g_3$, preserving $g_1$ and inversion $\mathcal{P}$. The spectrum of the full model $H_0+V$ is shown in  Fig. \ref{Magnetic_field_bulk}. 
According to the previous symmetry-related arguments, the breaking of time-reversal symmetry should leave the degeneracy of the eigenstates of $g_1$ symmetry operator at least at one point on $g_1$ invariant lines $k_y =0, \pm \pi$. However, in the present case, the degeneracy occurs at infinitely many points (i.e. on the entire segment  $X_2-M$ in  Fig. \ref{Magnetic_field_bulk}). 
This accidental degeneracy is, of course, a special feature of  a particular perturbation that may be lifted by a more general perturbation consistent with the symmetries. An additional accidental
degeneracy that arises occurs at the Weyl nodes $X_{1\pm}$ and $M_{1\pm}$ that emerge accidentally on the lines $k_x=\pm\pi$.
We now focus on these points, constituting a typical example of accidental Weyl nodes and explore the formation of the edge states in this case.
The position of the  Weyl points  can be easily found from  the bulk spectrum  (see the appendix \ref{A} for the details).
Its positions are:
\begin{align}
\label{weylpoints_mf_MX1}
X_{1\pm}=\left(\pi,\pm \arcsin \left[B_x/t^{\rm SO} \right] \right) \\
M_{\pm}=\left(\pi,\mp \left(\pi -\arcsin [B_x/t^{\rm SO}] \right)\right).
\end{align}
The split of the Weyl points $X_{1\pm}$, $M_{\pm}$ is controlled  by the ratio of the value of magnetic field and the amplitude of spin-orbit interaction.   The distance between the Weyl points therefore scales linearly with the magnetic field
\begin{equation}
\label{weyl_distance_mf_pi}
 \Delta k_y =\frac{2B_x}{t^{\rm SO}}.
 \end{equation}
 The Weyl points $X_{2\pm}$ on the line $k_x=0$ for $  t^{\rm SO}/t < 1$ and $B_x/ t^{\rm SO}<1$ are located at:
\begin{equation}
\label{weylpoints_mf_MX2}
\begin{split}
X_{2+}=\left(0,-\pi+\frac{B_x}{\sqrt{t^2+(t^{\rm SO})^2}}\right) \\
X_{2-}=\left(0,\pi-\frac{B_x}{\sqrt{t^2+(t^{\rm SO})^2}}\right) .
\end{split}
\end{equation}
For $t \gg t_{SO}$ the split between $X_{2\pm}$ is parametrically smaller then the distance between the Weyl points $X_{1-}$ and $X_{1+}$ or $M_+$ and $M_-$:
\begin{equation}
\label{weyl_distance_mf_0}
 \Delta k_y =\frac{2B_x}{\sqrt{t^2+(t^{\rm SO})^2}}.
 \end{equation}
 
\begin{figure}
\centering
\includegraphics[width=\linewidth]{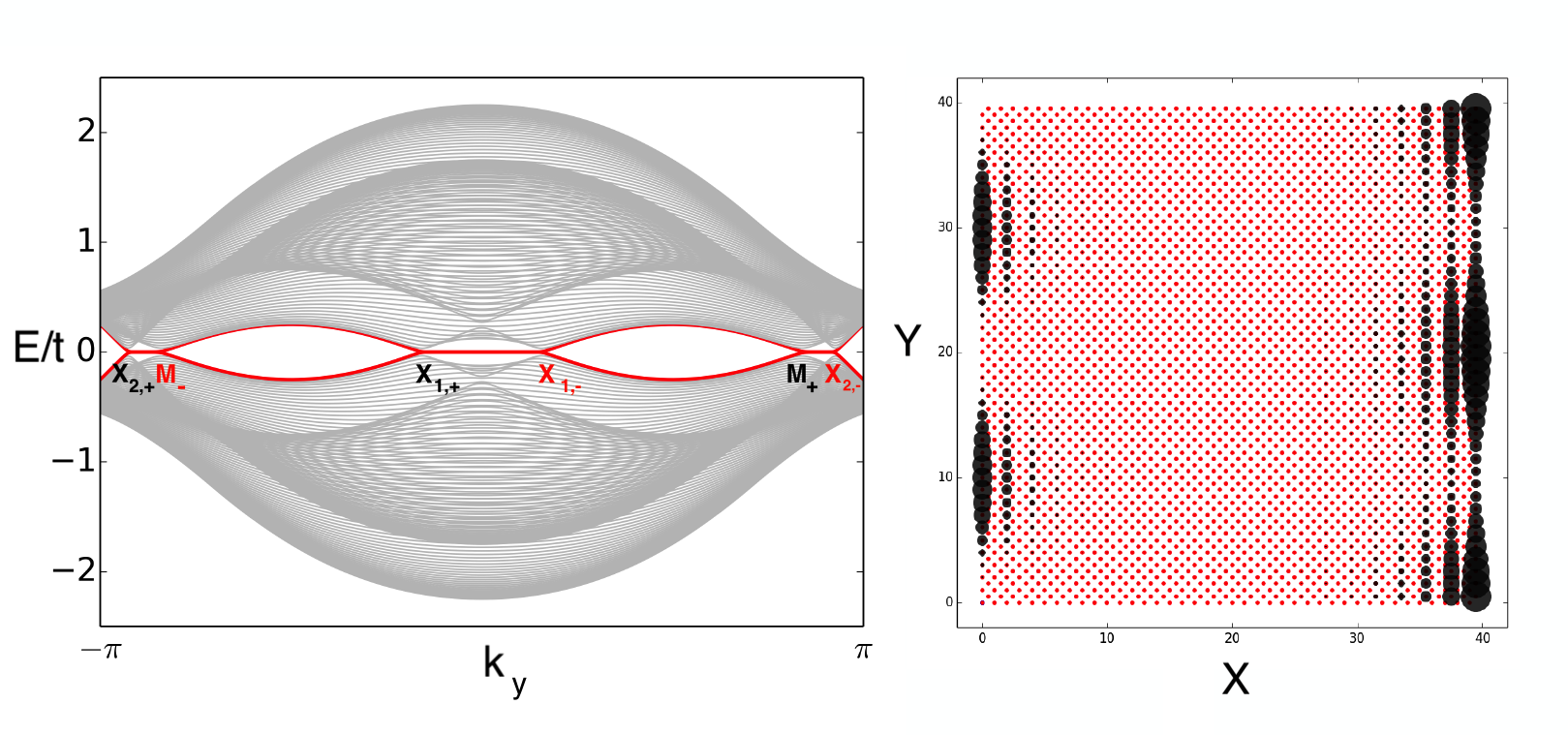}
\caption{\label{Magnetic_field_edge}
a) Spectrum of the model $H_0+V$ (magnetic field) for the ribbon with the width $L_x=60$, red line represents the band with the dispersionless edge states. b) Distribution of the amplitude of the wavefunction corresponding to the spin up component of the spinor in the ribbon with the width $L_x=40$}
\end{figure}

We now turn to a finite geometry and compute the spectrum of a  two-dimensional ribbon, 
which is infinite in $\hat{y}$ direction and has a finite width $L_x$ in $\hat{x}$ direction. Note, that in this setup the projections of the Weyl points located around the $M$ and $X_{1,2}$ points, which are associated with an opposite Chern number, do not overlap. 
The spectrum of the ribbon with $L_x=60$ is shown in the Fig. \ref{Magnetic_field_edge}. 
The edge states  connecting  pairs of Weyl points with  opposite topological charge are distinctly present.
The confinement in the real space depends  on a  direction of the spin, so that the right edge corresponds to electrons that are mostly in the up and the left edge to the electrons that  are mostly in the down spin state (see the Fig. \ref{Magnetic_field_edge}(b)).

\begin{figure}
\centering
\subfigure[]{ \includegraphics[width=0.7\linewidth]{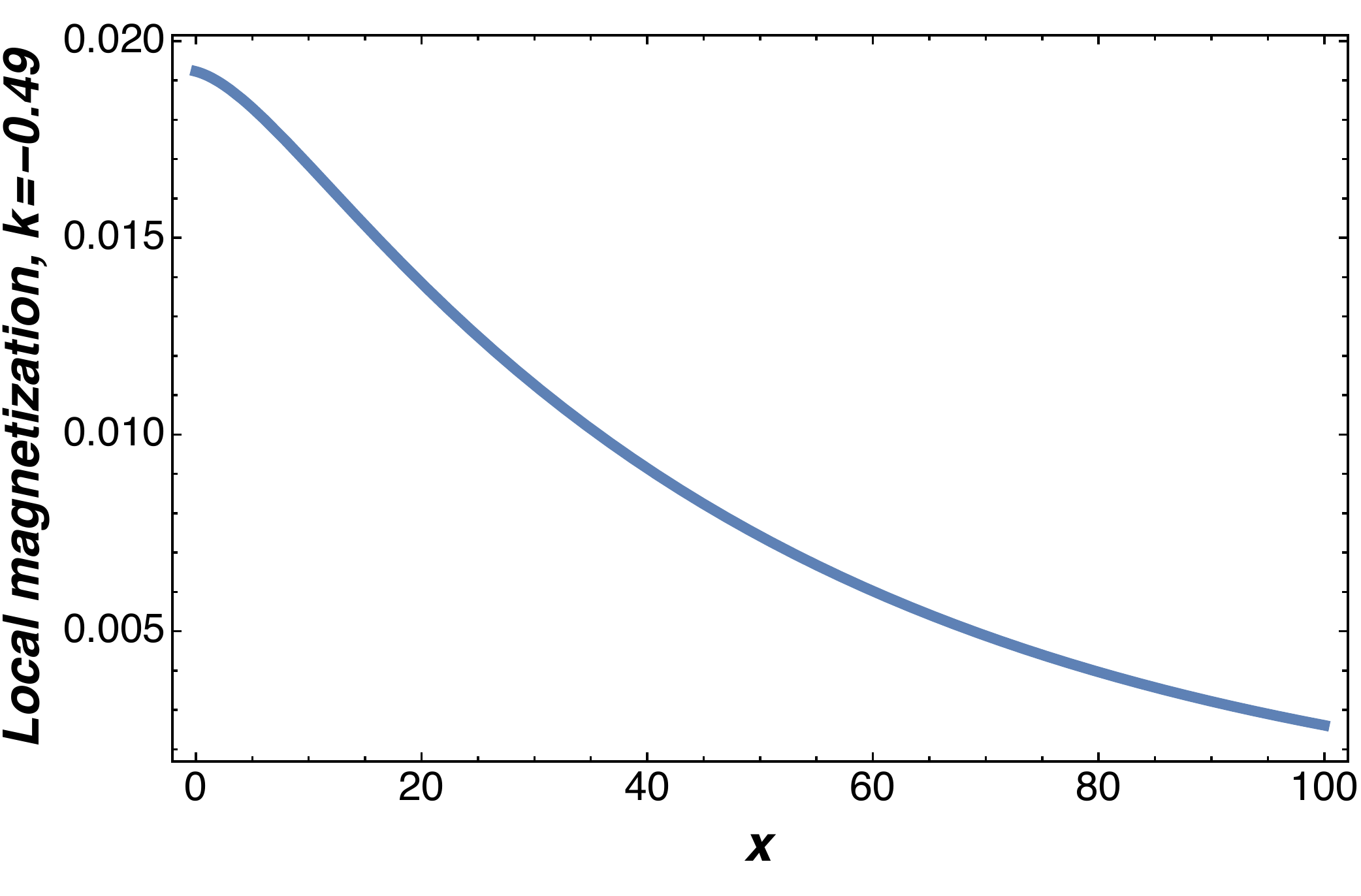}} \\
\subfigure[]{
\includegraphics[width=0.7\linewidth]{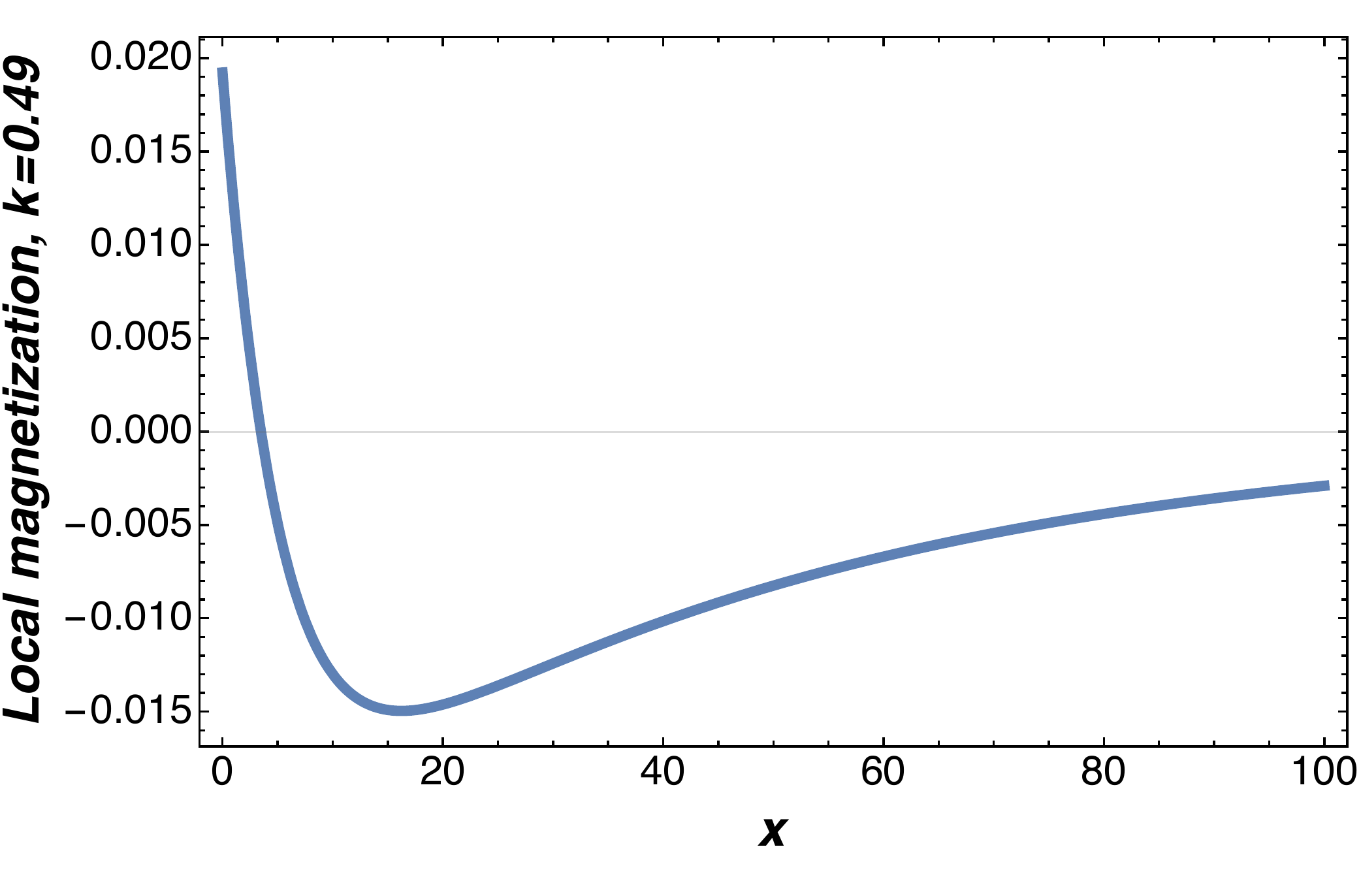}} \\
\subfigure[]{
\includegraphics[width=0.7\linewidth]{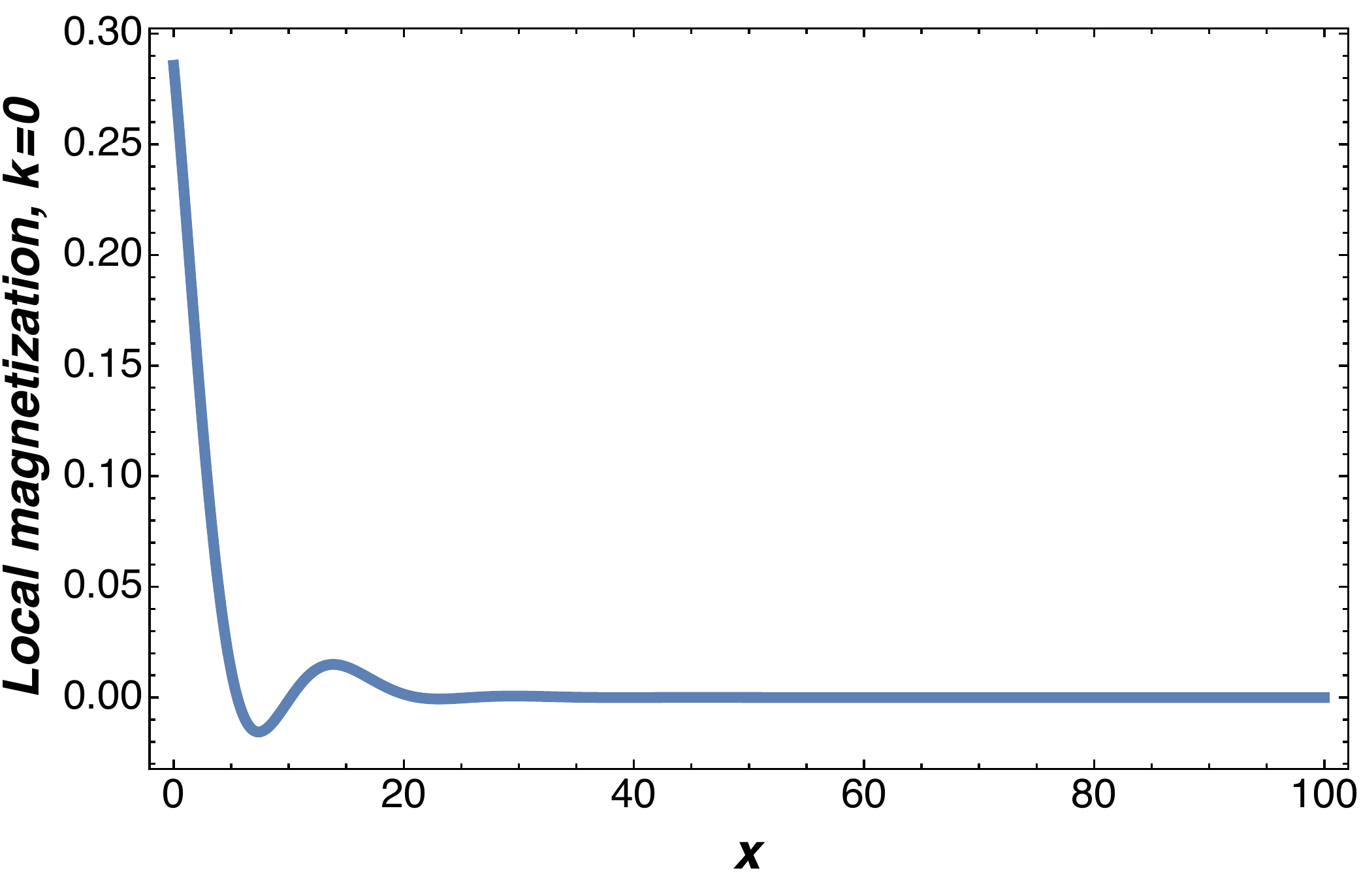}} 
\caption{ \label{Magnetization_local} Local magnetization of the semi-infinite system in magnetic field as a function of momentum $k_y$ along the edge, $t=2, t^{\rm SO}=1, B_x=0.5$. The panels  are drawn for  (a) $k_y=-0.49$, (b) $k_y=0.49$, (c) $k_y=0$. }
\end{figure}

This effect of spin polarization can be understood at the analytical level. To demonstrate that we study the magnetization density of the edge states for the model in magnetic field in the vicinity of the points $X_{1,\pm}$ on the half-plane (see Appendix B), which is defined as follows: 
\begin{equation}
\label{magnetization_density}
\langle M_z \rangle =\langle \Psi(x,k_y) |\sigma_z| \Psi(x,k_y) \rangle =\rho_{\uparrow}(x,k_y)-\rho_{\downarrow}(x,k_y)
\end{equation}
Whereas the magnetization of the edge state with the momentum $k_y$ is: 
\begin{equation}
\label{magnetization_total}
M(k_y)=\int_{0}^{\infty} dx \left[\rho_{\uparrow}(x,k_y)-\rho_{\downarrow}(x,k_y) \right]
\end{equation}
We see that the magnetization changes its sign when one moves from one Weyl point to another. So that in the vicinity of the Weyl point $X_{1,+} \simeq \left(\pi,-\frac{B_x}{t^{\rm SO}}\right)$ the state is almost polarized, $\rho_{\uparrow}(x,k_y) \rightarrow 1, \rho_{\downarrow}(x,k_y) \rightarrow 0$, and in the vicinity of $X_{1,-} \simeq \left(\pi,\frac{B_x}{t^{\rm SO}}\right)$ the polarization changes the sign. 
\begin{figure}
\centering
\includegraphics[width=\linewidth]{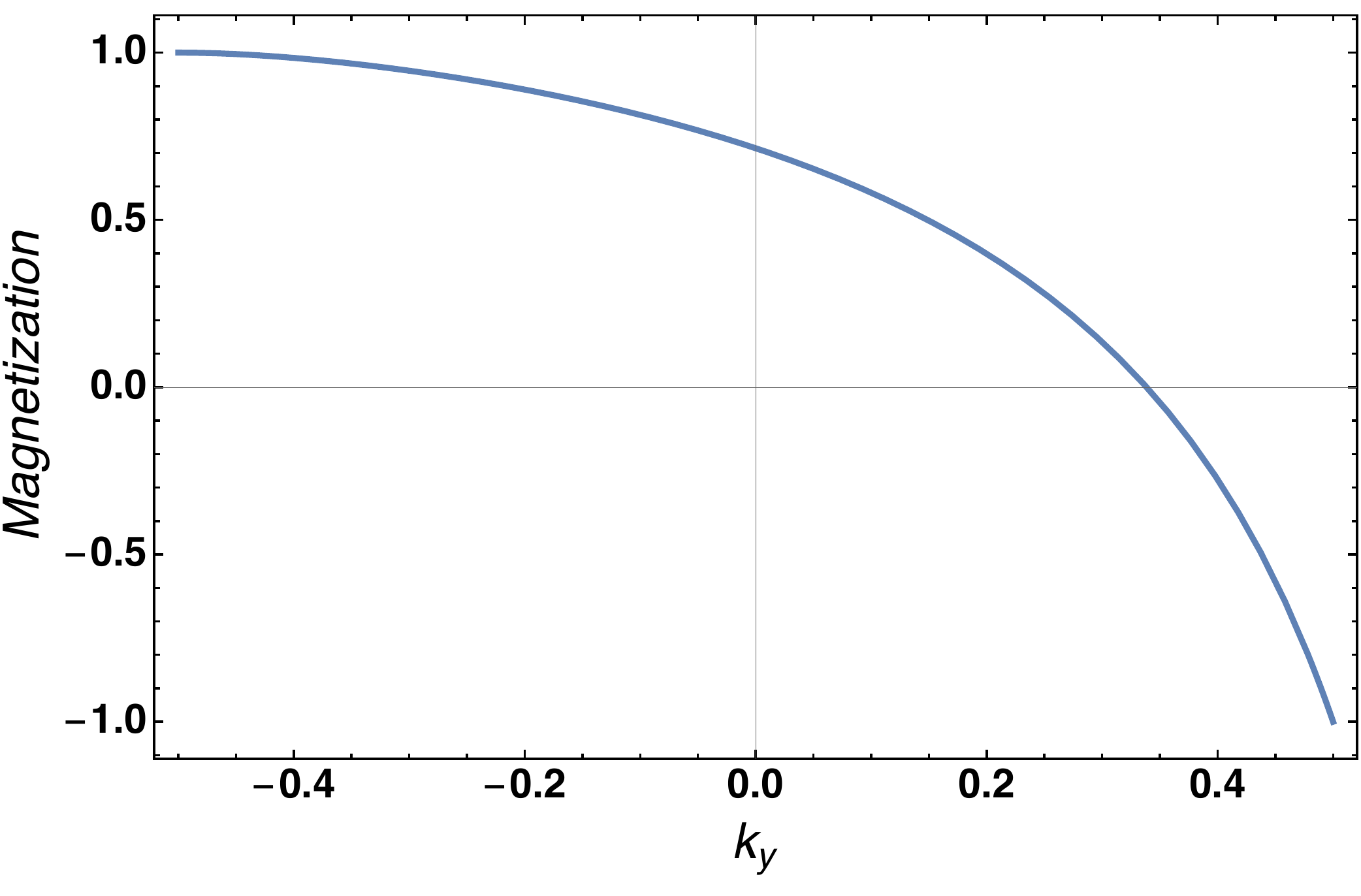}
\caption{\label{Magnetization_global} Magnetization of the semi-infinite system in magnetic field as a function of momentum $k_y$ along the edge, $t=2, t^{\rm SO}=1, B_x=0.5$}
\end{figure}
Distribution of the local magnetization as a function of coordinate is shown on the Fig. \ref{Magnetization_local}.  It demonstrates different behaviour in the vicinity of the different Weyl points and at $k=0$. Analytical formulas that describe this behavior are derived in the Appendix B (see Eqs.\ (\ref{magnetization_integrated_X1p})-(\ref{magnetization_local0}) there).

\section{Summary and Discussion}
We studied the formation of edge states in two-dimensional ribbons with non-symmorphic symmetries. 
By lowering the symmetries one may either open a gap or split the Dirac point into two Weyl nodes.
If at least one of non-symmorphic symmetries is preserved, the Weyl nodes appear on the symmetry invariant line in the  $k$ space. 
If non-symmorphic symmetries are broken the  outcome is undetermined. In a generic case, the spectrum is gapped  
however, for some perturbation  accidental  Weyl nodes may arise.

The type and the strength of the perturbation control  the type of emergent Weyl nodes and the composition of the wave function near these points (in terms of sublattices, spins, etc.).
It also determines the properties of the edge states  emerging at the boundaries of the ribbon.

In this work, we focused on two simple cases.
We applied a stress that displaced atoms in one direction, assuming that the atoms have magnetic dipoles aligned in the perpendicular direction. 
Since this perturbation breaks inversion symmetry but preserves one of the non-symmorphic symmetries, the  Weyl nodes are positioned on the corresponding invariant line.
Consequently, flat bands form in $k$ -space between the points that are projections of the nodes on the boundary. 

An accidental type of the Weyl nodes can be obtained by applying in-plane magnetic field.
The position of edge states is once again  consistent with the projection argument.
The spin structure of the Weyl nodes results in spin-polarized edge states, where the direction of polarization of the edge states is controlled by the sign of the magnetic field. The spatial distribution of the polarization depends on the strength of the magnetic field and on the value of momentum along the edge. 

While the calculations, that were performed, are specific to this model, some of the emergent features are expected to be general.
In particular, for a generic model where Weyl nodes  are separated from the bulk by a soft gap (i.e. there are no bulk excitations at these energies) excitations near the Weyl node will give rise to  edge states.
Therefore the problem in two dimensions is equivalent to a "deformed graphene" model, e.g. Eq. (\ref{blockhamiltonian_pert}), discussed above and the properties of the edge states can be determined within an approximate Dirac type Hamiltonian.

In the opposite case, where the excitation near the Weyl node coexists  with the regular bulk excitations 
the sample boundary may lead to the hybridization of those modes.
This  may (or may not) lead to the destruction of the edge states, but in any case, will  change their properties.
The precise mathematical criteria for this transition is, to the best of our knowledge,  yet to be constructed.

While the computation of edge states and their composition for realistic materials requires serious numerical analysis, 
the possibility to control the Weyl nodes by applying various symmetry breaking deformation to the Dirac semimetal constitutes a promising research route.
Choosing the type of deformation allows one to control the composition of the Weyl nodes, and the corresponding  edge states,
their dispersion and polarisation. 
This tunability  allows to utilize these materials as electromechanical sensors,
via their topological properties and responses, and in particular by measuring electric (and spin) current flowing along their edges.

\section{Acknowledgements}
We would like to thank H. Fertig, I. Gornyi and  P. Ostrovsky   for useful discussions.
D. G. was supported by ISF (grant 584/14)
and Israeli Ministry of Science, Technology and Space. D. M. acknowledges support from  the Israel Science Foundation (Grant No. 737/14) and from the People Programme (Marie Curie Actions) of the European Union’s Seventh Framework Programme (FP7/2007- 2013) under REA grant agreement No. 631064.
 \begin{appendices}
 \section{Spectrum on the $g_1$ invariant line for the model in magnetic field}
 \label{A}
In the presence of external in-plane magnetic field (\ref{magn_field}), the bulk spectrum of the system on $g_1$ invariant lines $k_x=\pm \pi$ is given by 
\begin{equation}
 \label{spectrum_magneticfield_pi}
\begin{split}
\epsilon_{1,2}(k_x=\pm\pi)= \pm t^{\rm SO} \left[ \frac{B_x}{t^{\rm SO}}+\sin k_y \right] \\
\epsilon_{3,4}(k_x=\pm \pi)= \pm t^{\rm SO} \left[ \frac{B_x}{t^{\rm SO}}-\sin k_y \right],
\end{split} 
 \end{equation}
and on the line $k_x=0$ by: 
\begin{small} 
 \begin{equation}
 \label{spectrum_magneticfield_0}
\begin{split}
\epsilon_{1,2}(k_x=0)= \pm t \left[ \frac{B_x}{t}+\sqrt{2}\cos \frac{k_y}{2}\sqrt{2+\frac{(t^{\rm SO})^2}{t^2}(1-\cos k_y)} \right] \\
\epsilon_{3,4}(k_x=0)= \pm t\left[ \frac{B_x}{t}-\sqrt{2}\cos \frac{k_y}{2}\sqrt{2+\frac{(t^{\rm SO})^2}{t^2}(1-\cos k_y)} \right] 
\end{split} 
 \end{equation}
 \end{small}
 Analyzing Eqs.(\ref{spectrum_magneticfield_pi}), (\ref{spectrum_magneticfield_0}), one finds the position of the Weyl nodes $X_{2,\pm}$  and $M_{\pm}$, see Eqs. (\ref{weylpoints_mf_MX1}),(\ref{weylpoints_mf_MX2}).
 
\section{Magnetization of the edge states}
\label{B} 
 Here we derive analytically the polarization of the edge states for the model in magnetic field. Consider the model $H_0+V$ in the vicinity of $(k_x,k_y)=(\pm \pi,0)$.
In the  linear order  in $k_x$, $k_y$ the Hamiltonian reads
\begin{equation}
\label{linearized_model}
H^{\text{lin}}= t^{\rm SO}\tau_z (\boldsymbol{\sigma}\times \mathbf{k})_z +B_x\sigma_x+tk_x \tau_x
\end{equation}
To account for a finite geometry we replace
$k_x \rightarrow -i \frac{\partial}{\partial x}$. 
The spin and sublattice  components can be combined into a spinor  $\Psi (k_y, x)=(\Psi^A, \Psi^B)$, where  we defined $\Psi^A=(\Psi^A_{\uparrow}, \Psi^A_{\downarrow})$ and $\Psi^B=(\Psi^B_{\uparrow}, \Psi^B_{\downarrow})$.
In terms of the spinor the Schr{\"o}dinger equation reads: 
\begin{small}
\begin{equation}
\label{equations}
\begin{cases}
\left[t^{\rm SO}(\sigma_x k_y+i\sigma_y\partial_x)+B_x\sigma_x \right] \Psi^A_x -it\partial_x \Psi_x^B=\epsilon \Psi_x^A \\
\left[t^{\rm SO}(-\sigma_x k_y-i\sigma_y\partial_x)  +B_x\sigma_x \right] \Psi_x^B -it\partial_x \Psi_x^A=\epsilon \Psi_x^B,
\end{cases}
\end{equation}
\end{small}
Where for the brevity of notations we denoted $\Psi_x^{A/B}=\Psi^{A/B}(x,k_y)$. 

In order to demonstrate the effect of polarization of the edge states, consider the problem (\ref{equations}) on the semi-infinite plane $x>0$  (or very wide ribbon $L_x \gg 1$) with the boundary conditions $\Psi^B(k_y, x=0)=0$. Solving this system of differential equations, we get the following zero-energy solution: 
\begin{equation}
\label{solution}
\begin{cases}
\Psi^A_{\uparrow}=0 \\
\Psi^A_{\downarrow}=C_1 e^{-k_1x}+C_2e^{-k_2x} \\
\Psi^B_{\uparrow}=C_3 \left(e^{-k_1x}-e^{-k_2x}\right) \\
\Psi^B_{\downarrow}=0,
\end{cases}
\end{equation}
where $C_1, C_2, C_3$ are functions of  the parameters $t, t^{\rm SO}, B_x$ and  $k_y$. The penetration depths of magnetization are 
\begin{equation}
\label{decaying_length}
\begin{split}
\frac{1}{\lambda_1}=k_1=\frac{B_x t^{\rm SO}+\sqrt{\left[t^2+(t^{\rm SO})^2\right] (t^{\rm SO})^2 k_y^2-t^2B_x^2}}{t^2+(t^{\rm SO})^2} \\
\frac{1}{\lambda_2}=k_2=\frac{B_xt^{\rm SO} -\sqrt{\left[t^2+(t^{\rm SO})^2\right] (t^{\rm SO})^2 k_y^2-t^2B_x^2}}{t^2+(t^{\rm SO})^2}.
\end{split}
\end{equation}
Notice that for $B_x>0$ the exponentially decaying solution (i.e., $k_1, k_2>0$) exists, provided  $|k_y|<B_x/ t^{\rm SO}$. This means that the momentum  $k_y$  vary between the two Weyl  nodes $X_{1,\pm} =(\pi,\pm\frac{B_x}{t^{\rm SO}})$. 
This result was obtained within the  linearized model but is expected to hold in general.  
Now we calculate the local and total magnetization of the model, defined in (\ref{magnetization_density}), (\ref{magnetization_total}). 
Around the Weyl point $X_{1,+} \simeq \left(\pi,-\frac{B_x}{t^{\rm SO}}\right)$ the total magnetization is described by: 
\begin{equation}
\label{magnetization_integrated_X1p}
M(k_y \simeq X_{1,+})=1-\frac{t^2}{8B_x^2}(\delta k_y)^2,
\end{equation}
where $\delta k_y=X_{1,+}-k_y$. Around the Weyl point $X_{1,-}$ the magnetization changes the sign, 
\begin{equation}
M(k_y \simeq X_{1,-})=-1+\frac{4t^{\rm SO}(t^2+(t^{\rm SO})^2)}{B_xt^2} \delta k_y,
\end{equation}
where $\delta k_y=X_{1,-}-k_y$. 
The local magnetization also shows different behaviour depending on the momentum $k_y$.  Close to the point $X_{1,+}$ one has: 
\begin{equation}
\label{magnetization_localX1p}
M_{X_{1,+}}(x) =2\delta k_y +\left[\frac{t^2\left(1-2e^{-\frac{x}{\lambda_M}}\right)}{B_xt^{\rm SO}} -4x \right]  \delta k_y^2
\end{equation}
where $\lambda_M$ is defined as:
\begin{equation}
\label{lambda_magnet}
\lambda_M=\frac{t^2+(t^{\rm SO})^2}{2B_xt^{\rm SO}},
\end{equation}
and the magnetization around the point $X_{1,-}$ is:
\begin{equation}
\label{magnetization_localX1m}
M_{X_{1,-}}(x) = e^{-\frac{x}{\lambda_M}}\left[4 \sinh^2 \frac{x }{2\lambda_M}-\frac{8(t^{\rm SO})^2}{t^2} e^{-\frac{x}{\lambda_M}} \right]\delta k_y 
\end{equation}

However, around the point $k_y=0$ the local magnetization oscillates. 
\begin{equation}
\label{magnetization_local0}
M_{k_y=0}(x) = \frac{2B_x t^{\rm SO}e^{-\frac{x }{\lambda_M}}}{5t^2+8(t^{\rm SO})^2}\left(3+5\cos \frac{x}{\lambda_M} \right)
\end{equation}

Our results we derived assuming that $B_x>0$. 
The magnetization for the negative field acquires an opposite sign.
 This  follows from  (\ref{equations}), since  the states  $\Psi^A_{\downarrow} \rightarrow \Psi^A_{\uparrow}$ and $\Psi^B_{\uparrow} \rightarrow \Psi^B_{\downarrow}$
 under reversal of a magnetic field ($B_x \rightarrow -B_x$).
 Also notice that one can perform the similar analysis of the magnetization around the other Weyl points, $M_{\pm}$ and $X_{2\pm}$, and the qualitatively similar results for that case can be obtained by the replacement $t \rightarrow \frac{tk_y}{2}$. 
 \end{appendices}


\begin{thebibliography}{10}
\bibitem{KaneYoung}

S. M. Young and C. L. Kane, "Dirac semimetals in two dimensions",  Phys. Rev. Lett. {\bf 115}, 126803 (2015).

\bibitem{Binghai1}
Binghai Yan and Claudia Felser, "Topological Materials: Weyl Semimetals", Annual Review of Condensed Matter Physics {\bf 8}, 337-354 (2017).

\bibitem{Xu}
Su Yang-Xu et al., "Discovery of a Weyl Fermion semimetal and topological Fermi arcs", Science {\bf 349}, 613 (2015).

\bibitem{Lv}
B.Q. Lv et al., "Experimental Discovery of Weyl Semimetal TaAs", Phys. Rev. X {\bf 5}, 031013 (2015).

\bibitem{Lu}
L. Lu, Z. Wang, D. Ye, L. Ran, L. Fu, J. D. Joannopoulos and M.  Solja{\v c}i{\' c}, "Experimental observation of Weyl points" , Science {\bf 349}, 622 (2015).

\bibitem{Burkov}A.A.  Burkov,  and L.  Balents, "Weyl semimetal in a topological insulator multilayer",  Phys. Rev. Lett. {\bf 107}, 127205 (2011).

\bibitem{Burkov2} 
A.A.  Burkov, "Chiral anomaly and transport in Weyl metals", Journal of Physics: Condensed Matter {\bf 27},11 (2015).

\bibitem{NielsenNinomiya}
H. Nielsen and M. Ninomiya, "The Adler-Bell-Jackiw anomaly and Weyl fermions in a crystal",  Phys. Lett. B {\bf 130}, 389 (1983).

\bibitem{Wan}
Xiangang Wan, Ari M. Turner, Ashvin Vishwanath and Sergey Y. Savrasov, "Topological semimetal and Fermi-arc surface states in the electronic structure of pyrochlore iridates", Phys. Rev. B {\bf 83}, 205101 (2011).


\bibitem{Singh}
Bahadur Singh, Ashutosh Sharma, H. Lin, M. Z. Hasan, R. Prasad, and A. Bansil, "Topological electronic structure and Weyl semimetal in the $\text{TlBiSe}_2$ class of semiconductors", Phys. Rev. B {\bf 86}, 115208 (2012).

\bibitem{Potter}
A.C. Potter, I. Kimchi,  and A. Vishwanath, "Quantum oscillations from surface Fermi arcs in Weyl and Dirac semimetals", Nat. Commun. {\bf 5}, 5161 (2014).

\bibitem{Huang}
Shin-Ming Huang et al., "A Weyl Fermion semimetal with surface Fermi arcs in the transition metal monopnictide TaAs class", Nat Commun. {\bf 6}, 7373 (2015). 

\bibitem{Solyanov}
A.A. Soluyanov et al., "Type-II Weyl semimetals", Nature {\bf 527}, 495–498 (2015).

\bibitem{Binghai2}
Yan Sun, Shu-Chun Wu, and Binghai Yan, "Topological surface states and Fermi arcs of the noncentrosymmetric Weyl semimetals TaAs, TaP, NbAs, and NbP", Phys. Rev. B {\bf 92}, 115428 (2015).

\bibitem{Binghai3}
L. X. Yang et al., "Weyl semimetal phase in the non-centrosymmetric compound TaAs", Nature Physics {\bf 11}, 728–732 (2015).

\bibitem{Weng}
Hongming Weng, Chen Fang, Zhong Fang, B. Andrei Bernevig, and Xi Dai, "Weyl Semimetal Phase in Noncentrosymmetric Transition-Metal Monophosphides", Phys. Rev. X {\bf 5}, 011029 (2015).

\bibitem{Batabyal}
Rajib Batabyal et al., "Visualizing weakly bound surface Fermi arcs and their correspondence to bulk Weyl fermions", Science Advances, {\bf 2}, 8 (2016).

\bibitem{Deng}
Ke Deng et al., "Experimental observation of topological Fermi arcs in type-II Weyl semimetal $\text{MoTe}_2$", Nature Physics, {\bf 12}, 1105â1110 (2016).

\bibitem{Li}
Feng Li, Xueqin Huang, Jiuyang Lu, Jiahong Ma and Zhengyou Liu, "Weyl points and Fermi arcs in a chiral phononic crystal", Nature Physics {\bf 14}, 30-34 (2018).

\bibitem{Jiang}
J.Jiang et al., "Signature of type-II Weyl semimetal phase in $\text{MoTe}_2$", Nat. Commun. {\bf 8}, 13973 (2017).

\bibitem{Armitage}
N.P. Armitage, E.J. Mele, and A. Vishwanath, "Weyl and Dirac semimetals in three-dimensional solids", Rev. Mod. Phys. {\bf 90}, 015001 (2018).

\bibitem{Symmetries_in_solids}
C.J.Bradley and A.P. Cracknell, "The mathematical theory of symmetry in solids", Clarendon Press, Oxford (1972). 

\bibitem{FujitaJPSJ1996}
Mitsutaka Fujita, Katsunori Wakabayashi, Kyoko Nakada, and Koichi Kusakabe, "Peculiar Localized State at Zigzag Graphite Edge", J. Phys. Soc. Jpn. {\bf 65}, 1920 (1996).

\bibitem{NakadaPRB1996}
Kyoko Nakada, Mitsutaka Fujita, Gene Dresselhaus, and Mildred S. Dresselhaus, "Edge state in graphene ribbons: Nanometer size effect and edge shape dependence", Phys. Rev. B {\bf 54}, 17954 (1996).

\bibitem{WakabayashiPRB1999}
Katsunori Wakabayashi, Mitsutaka Fujita, Hiroshi Ajiki, and Manfred Sigrist,"Electronic and magnetic properties of nanographite ribbons", Phys. Rev. B {\bf 59}, 8271  (1999).

\bibitem{Kane_Mele}
C. L. Kane and E. J. Mele, "Quantum Spin Hall Effect in Graphene", Phys. Rev. Lett. {\bf{95}}, 226801(2005).

\bibitem{Kane_Fu}L. Fu and C. L. Kane, "Topological insulators with inversion symmetry", Phys. Rev  B {\bf 76}, 045302 (2007).

\bibitem{Yang}
Hua Tong Yang, "Strain-induced shift of Dirac points and the pseudo-magnetic field in graphene", Journal of Physics: Condensed Matter, {\bf{23}}, 50 (2011).

\bibitem{Ryu}
S. Ryu and Y.Hatsugai, "Topological origin of zero-energy edge states in particle-hole symmetric systems", Phys.Rev.Lett. {\bf{89}}, 077002 (2002).


\bibitem{Levitov}  O. Shtanko and  L. Levitov, "Robustness and Universality of Surface States in Dirac Materials", Proceedings of the National Academy of Sciences, 201722663 (2018).

\bibitem{Kharitonov}
M. Kharitonov, J.-B. Mayer, E.M. Hankiewicz, "Universality and Stability of the Edge States of Chiral-Symmetric Topological Semimetals and Surface States of the Luttinger Semimetal",  Phys. Rev. Lett. {\bf 119}, 266402 (2017).

\bibitem{CastroNeto}
A. H. Castro Neto, F. Guinea, N. M. R. Peres, K. S. Novoselov, and A. K. Geim, "The electronic properties of graphene",  Rev. Mod. Phys. {\bf 81}, 109 (2009).

\bibitem{NielsenNinomya2}
H. Nielsen and M. Ninomiya, "Absence of neutrinos on a lattice (I). Proof by homotopy theory", Nuclear Physics B {\bf 185}, 20-40 (1981).


\end{thebibliography}
\end{document}